# FORMATION MECHANISMS OF SINGLE-PULSE AND SECONDARY ECHOES IN SYSTEMS WITH A LARGE INHOMOGENEOUS BROADENING OF NMR LINES

*J.G.Chigvinadze, G.I.Mamniashvili, Yu.G.Sharimanov*

E.Andronikashvili Institute of Physics of Georgian Academy of Sciences

**Abstract**

In the frames of Mims transformation matrix method the equation for nuclear magnetization are obtained describing the dynamics of nuclear spin-systems with a strong Larmor and Rabi inhomogeneous broadenings of NMR line in conditions of their unequlibrium earlier obtained by the statistical tensors method. As example, properties of proton single-pulse echo and its secondary signals in probe material (silicon oil) coated on the surface of high-$T_c$ superconducting-oxides powder and in metallic hydride are presented.

The single-pulse echo (SPE) is a resonance response of the inhomogeneously broadened nuclear spin-system to the application of a solitary radiofrequency (RF) pulse arising at a time moment approximately equal to the pulse duration $\tau$ after its termination. Though SPE was discovered by Bloom in 1955 for protons in water placed in an inhomogeneous magnetic field, the mechanism of SPE formation is not yet so clear-out as for the classical Hahn two-pulse echo (TPE) and it continues to attract researchers attention [1].

The point is that the theoretical models based exclusively on strong Larmor inhomogeneous broadening (LIB) do not agree with theoretically obseved signals, but instead result in the formation of oscillatory free induction decays (OFID's) [1].

SPE formation mechanisms could be conditionally subdivided into two classes: the first one is so called edge-type mechanisms where RF pulse edges act like RF pulses in the TPE method such as the distortion mechanism [1], and mechanism connected with the consideration of spectral densities of sufficiently step RF pulse edges [2], and the second class is mechanisms of internal nature due to particular type nonlinearities in the dynamics of spin-systems, as example, connected with a strong dynamic frequency shift of NMR frequency, or with a nonlinear dynamics of nuclear spins due to the simultaneous presence of large Larmor and Rabi inhomogeneous broadenings of NMR line [1].

In this work we consider in more details so-called multipulse mechanism of SPE formation, presented in work [1], for systems with the both types of frequency inhomogeneities of NMR lines. The important example of such systems are nuclei arranged in the domain walls (DW) of multidomain magnets and both in normal metals due to the metallic skin effect, and in normal cores of Abrikosov vortices in type II superconductors. Earlier in [3] we have investigated properties of the SPE formation In lithium ferrite. It was established that its properties are sharply differ from SPE properties in hexagonal cobalt, where it is formed by the distortion mechanism. Therefore the conclusion was made on a possible effectivity of the internal mechanism of SPE formation in lithium ferrite. But its concrete mechanism was not finally established.

Later on the effectiveness of multipulse mechanism of the SPE formation was experimentally established in this magnetics. Moreover, the secondary echo signals of SPE and the two-pulse echo (TPE) were also formed by this mechanism.

In [1] it was shown that multipulse mechanism of SPE formation was effective in some multidomain ferromagnets like Fe and FeV. From this point of view further theoretical and experimental investigations of SPE multipulse mechanism formation in systems with large Larmor and Rabi inhomogeneous broadenings of NMR lines are of practical interest.

In work [1] using the formalism of statistical tensors the theoretical investigation of SPE and its secondary echo signals formation mechanism was made allowing for both large Larmor and Rabi inhomogeneous broadenings of NMR line when repetition period of RF pulses T obeys the inequality $T_3 \ll T_2 < T < T_1$, where $T_1$, $T_2$, $T_3$ were characteristic NMR relaxation parameters, therefore a spin-system was in unequilibrium state before the application of exciting RF pulse, and thereby only a longitudinal component of the nuclear magnetization was important before RF pulse. It was shown that a dephasing of nuclear spin-system was accumulated during n-time pulse excitations and restored within a time interval elapsing from the trailing edge of the last "counting" [(n+1) th] pulse in the multipulse train. This resulted in the SPE formation and also its secondary signals at time moments which were multiples of the RF duration after a termination of the "counting" RF pulse.

Let us show further a simple classical derivation of the equations describing the nuclear spin-system dynamics in the investigated case, in the frames of usual classical approach by solving Bloch equations or by the equivalent Mim's transformation matrix method [4]. We will use the last one as the most visual from the experimental point of view.

Let us consider a case when a local static field $H_n$ is directed along $Z$ axis, and a PF field is along X axis of RSC [5,6]. $\mathbf{H}_{eff}$ modulus in RSC could be expressed by:

$$H_{efff} = \frac{1}{\gamma}\sqrt{\Delta\omega_j^2 + \omega_1^2} = \frac{\varpi_1}{\gamma_n}\sqrt{a^2 + x^2} \qquad (1)$$

Here $x = \Delta\omega_j / \varpi_1$, where $\Delta\omega_j = \omega_j - \omega_o$ isochromate frequency $a = \eta/\bar{\eta}$ (or $a = \omega_1 / \varpi_1$), where $\eta$ is the RF field gain factor and $\bar{\eta}$ its mean value; $\varpi = \bar{\eta}\omega_1^{APPL}$ - is a mean value of RF amplitude in frequency units; $\omega_1 = \eta\omega_1^{APPL}$ is Rabi frequency of applied RF field. Let us introduce, besides this [1], correspondingly, following designations for a mean value of a pulse area $y = \varpi_1 \Delta t$, where $\Delta t$ is a RF pulse duration $b = \varpi \tau$ is a characteristic of a time interval following a pulsed excitation and is measured from the trailing edge of RF pulse; $\omega_o$ is designation for a center of the resonance line, and $\omega_j$ is a frequency of $j$-th isochromate, correspondingly. The transformation matrix describing a rotation of the magnetization vector around $\mathbf{H}_{eff}$ is [4]: $\bar{m} = (\bar{m}_x; \bar{m}_y; \bar{m}_z)$

$$(R) = \begin{bmatrix} S_\phi^2 + C_\phi^2 C_\theta & -C_\phi S_\theta & S_\phi C_\phi (1 - C_\theta) \\ C_\phi S_\theta & C_\theta & -S_\phi S_\theta \\ S_\phi C_\phi (1 - C_\theta) & S_\phi S_\theta & C_\phi^2 + S_\phi^2 C_\theta \end{bmatrix} \qquad (2)$$

$C_\phi$, $S_\phi$, $C_\theta$ and $S_\theta$ stand for $\cos\phi$, $\sin\phi$, $\cos\theta$ and $\sin\theta$, and $\Psi = tg^{-1}\left(\dfrac{\omega_1}{\Delta\omega_j}\right)$ is the angle between the effective field $\mathbf{H}_{\text{eff}}$ and $Z$ axis; $\theta$ is the angle by which the magnetization turns about the effective field $\mathbf{H}_{\text{eff}}$ during the pulse time $\Delta t$:

$$\theta = \nu \mathbf{H}_{\text{eff}} t, \quad \text{where } \mathbf{H}_{\text{eff}} \text{ is given by (1)}.$$

Let us consider firstly the case of single-pulse excitation. Let

$$X_j = m_{xj}/m; \quad Y_j = m_{yj}/m; \quad Z_j = m_{zj}/m \text{ and } \overline{\mu} = (X_j; Y_j; Z_j)$$

where $m$ is an equilibrium nuclear magnetization and at the equilibrium $\overline{\mu}_{eq} = (0; 0; 1)$.

It before the excitation by RF pulse a nuclear spin-system was at equilibrium conditions, and therefore $\overline{\mu}_{eq} = (0; 0; 1)$, then a result of RF pulse action is presented by $\overline{\mu} = (R)\overline{\mu}_{eq}$.

At the termination of RF pulse isochromates precess freely around $Z$ axis what is described by the matrix:

$$R_\varphi = \begin{pmatrix} C_\varphi & -S_\varphi & 0 \\ S_\varphi & C_\varphi & 0 \\ 0 & 0 & 1 \end{pmatrix},$$

where $\varphi = \Delta\omega_j \tau$ is the turning angle of isochromate around Z axis, and $\tau$ is a time elapsing from the trailing and of a pulse. Therefore, we have finally:

$$\overline{\mu}_1 = (R_\Phi)(R)\overline{\mu}_{eq} = \begin{pmatrix} C_\varphi S_\phi C_\phi (1-C_\theta) + S_\varphi S_\phi S_\theta \\ \\ S_\varphi S_\phi C_\phi (1-C_\theta) - C_\varphi S_\phi S_\theta \\ \\ C_\phi^2 + S_\phi^2 C_\theta \end{pmatrix},$$

or in the accepted designations:

$$\frac{m_x}{m} = \cos bx \frac{ax}{a^2+x^2}\left(1 - \cos y\sqrt{a^2+x^2}\right) + \sin bx \frac{a}{\sqrt{a^2+x^2}} \sin y\sqrt{a^2+x^2}$$

$$\frac{m_y}{m} = \sin bx \frac{ax}{a^2+x^2}\left(1 - \cos y\sqrt{a^2+x^2}\right) - \cos bx \frac{a}{\sqrt{a^2+x^2}} \sin y\sqrt{a^2+x^2}$$

$$\frac{m_z}{m} = 1 - \frac{a^2}{a^2+x^2}\left(1 - \cos y\sqrt{a^2+x^2}\right)$$

This expressions coincide with the correspondent ones obtained in [1] for the case of single-pulse excitation, and similar expressions [7] obtained by solving the system of Bloch equations for inhomogeneously broadened Hahn systems.

Let us find now the effect of n-time RF excitation in the frames of model [1], when before next RF pulse of train only the longitudinal component of nuclear magnetization remains. It is not difficult to prove by successive matrix multiplication that expressions for nuclear magnetization before final "counting" (n+1)-th pulse are:

$$\bar{\mu}_n = \left(C_\phi^2 + S_\phi^2 C_\theta\right)^n \bar{\mu}_{eq}, \text{ where } \bar{\mu}_{eq} = (0; 0; 1).$$

Then the result of excitation by the "counting" pulse and following free precession of magnetization is described by expressions:

$$\bar{\mu}_{n+1} = (R_\Phi)(R)\bar{\mu}_n = \left(C_\phi^2 + S_\phi^2 C_\theta\right)^n \begin{pmatrix} C_\varphi S_\phi C_\phi (1-C_\theta) + S_\varphi S_\phi S_\theta \\ \\ S_\varphi S_\phi C_\phi (1-C_\theta) - C_\varphi S_\phi S_\theta \\ \\ C_\phi^2 + S_\phi^2 C_\theta \end{pmatrix},$$

which is similar the one for single-pulse excitation but allowing for a new initial condition.

It follows from previous expression in accepted designation:

$$\frac{m_x}{m} = \left(1 - \frac{a^2}{a^2 + x^2}\left[1 - \cos y\sqrt{a^2 + x^2}\right]\right)^n$$
$$\left[\cos bx \frac{ax}{a^2 + x^2}\left(1 - \cos y\sqrt{a^2 + x^2}\right) + \sin bx \frac{a}{\sqrt{a^2 + x^2}} \sin y\sqrt{a^2 + x^2}\right]$$

$$\frac{m_y}{m} = \left(1 - \frac{a^2}{a^2 + x^2}\left[1 - \cos y\sqrt{a^2 + x^2}\right]\right)^n$$
$$\left[\sin bx \frac{ax}{a^2 + x^2}\left(1 - \cos y\sqrt{a^2 + x^2}\right) + \cos bx \frac{a}{\sqrt{a^2 + x^2}} \sin y\sqrt{a^2 + x^2}\right]$$

These expressions coincide with the ones obtained in [1] using the formalism of statistical tensors. The n-th degree multipulse has a simple physical meaning of a longitudinal nuclear magnetization created by n preliminary pulses of a multipulse train reflecting the memory of spin-system on the excitation. The expressions for SPE and its secondary echo signal amplitudes using similar expressions for nuclear magnetization vectors were already obtained in [1]. It is easy to prove that above considered approach could be immediately applied to the case of periodical two-pulse excitation, what is of interest for description of secondary echo signals in the investigated systems.

Let us know also [1] that the effect of SPE and its secondary echo signals formation is present for a large LIB in isolation but is stronger at the simultaneous presence of both frequency inhomogeneities as in the case of multidomain ferromagnet and type II superconductors.

Let us illustrate some of the above pointed dependences on the concrete examples having a practical interest.

Experimental results were obtained by the NMR spectrometer Bruker "Minispec p20" provided with the digital signal averager "Kawasaki Electronica" at room and liquid nitrogen temperatures.

Fig.1 shows the averager record of SPE and its secondary signals of protons in a liquid solution of $MnCl_2$ at the periodical excitation by a pulse train with the period T=4 ms. The longitudinal and transverse relaxation times are, correspondingly, $T_1$=86 ms and $T_2$=72 ms.

Let us consider in more details the SPE signal formation on the example of protons in the probe material (silicon oil (SO), Silicon KF96) coated on the surface of powdered sample of high-$T_c$ superconductor (HTSC) YBCO-(SO+YBCO) which is object similar one used in work [8] to study the effect of inhomogeneous broadening of NMR lines due to the formation of Abrikosov vortex lattice in HTSC.

Fig.2 shows the SPE record of investigated sample (SO+YBCO), and in Fig.3 its intensity dependence on the RF pulse period T at room temperature (a), and at liquid nitrogen temperature (b).

Let us note that at the given maximal RF pulse length of spectrometer (20 μs) for the observation of SPE signal one should introduce an artificial outer magnetic field inhomogeneity (with a help of additional iron plate [9]) to allow for condition (1)). At the same time at T=77 K (b), the SPE signal is observed in an homogeneous magnetic field, but the inhomogeneity of NMR line is caused by the effect of the AVL formation.

The character of dependence on T points on a comparatively large role of the multipulse mechanism in the SPE formation at low temperatures.

The SO concentration in a sample under investigation was chosen minimal for the inhancement of the vortex lattice effect [8].

To compare with, in Fig,4 (a) a record of TPE and its secondary echo signals is presented of SO+YBCO sample with larger concentration of coating material to obtain more intensive signals, but in fig, 4(b) the intensity dependences of TPE (1) and its secondary signal (2) on the period of two-pulse train T.

It is seen that dependences of SPE and secondary TPE signals on T have a similar character reflecting the significant contribution of multipulse mechanism in the SPE intensity. It is known that secondary TPE signals are formed by the multipulse mechanism in proton containing systems [10]

Vanadium hydride ($VH_{0.68}$) could be considered as one more example of systems possessing both types inhomogeneities. In this case the inhomogeneities are a result of the metallic skin effect. In Fig.5 the dependence of SPE signal on T at room temperature are presented. In this case its intensity practically not changed at increasing of T showing that contribution of the distorsion mechanism is significant in this material as it is in some metallic ferromagnets [1].

The analysis of obtained results shows that the SPE could be useful not only for a simple determination of the characteristic relaxation parameters of inhomogeneously broadened spin-systems, but could provide an interesting approach to the study of AVL dynamics using the SPE signal due to the effect of magnetic field inhomogeneity caused by the AVL formation

This allows one to use the SPE effect for the AVL stimulated dynamics study using pulsed and low frequency magnetic fields [11].

In conclusion, in frames of a simple classical approach using Mim's transformation matrix method, the equation for nuclear magnetization are obtained describing the dynamics of nuclear spin-systems with strong Larmor and Rabi inhomogeneous broadenings of NMR lines in conditions of their unequilibrium.

Properties of proton single-pulse echo and its secondary signals in probe material (silicon oil) coated on the surface of high-$T_c$ superconducting-oxides powder and in metallic hydride are presented.

Besides, it is experimentally shown that the single-pulse echo effect gives the opportunity to obtain valuable information on the inhomogeneous NMR broadening reflecting a character of microscopic distribution of magnetic field in such systems as superconductors, hydrides of metals and so on.

Acknowlegment

This work was supported by the International Science and Technology Centre through Project G-389.

**Figure captions**

Fig.1 Single-pulse echo (SPE) and its secondary signals in liquid solution MnCl$_2$

$\tau$=20μs, T=4ms, T$_1$=86ms, T$_2$=72ms.

Fig.2 SPE in silicon oil (SO) mixed with YBCO powder

Fig.3 SPE dependence on the period of single-pulse train T at room temperature (a) and at liquid nitrogen temperature (b)

Fig.4  a- Two-pulse echo (TPE) and its secondary signals in SO + YBCO;

 b- Intensity dependences of TPE (1), and its secondary signal (2) on the period two-pulse train T.

Marks show time position of RF pulses.

Fig.5 The dependence of SPE intensity on the period of single-pulse train T in vanadium hydride (VH$_{0.68}$) $\tau$=20μs,  T=300K.

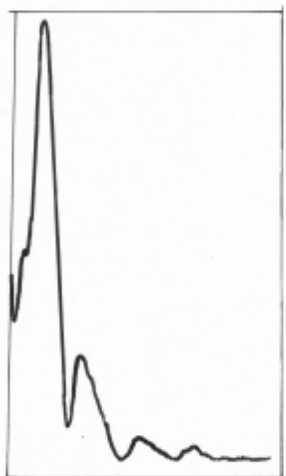
Fig.1

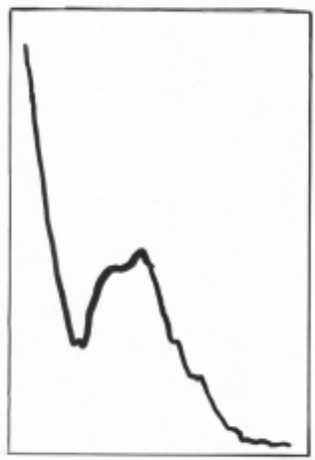
Fig.2

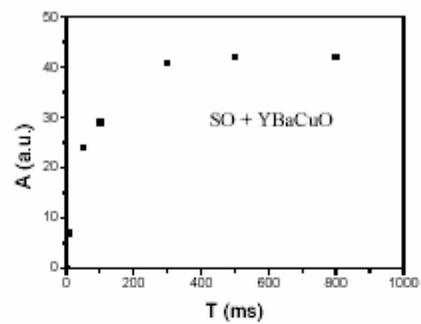
Fig.3a

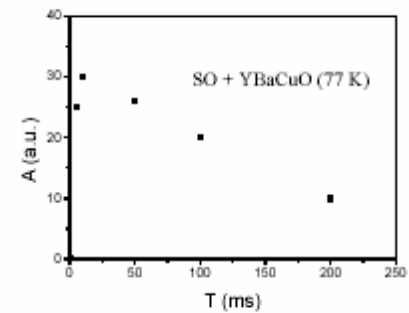
Fig.3b

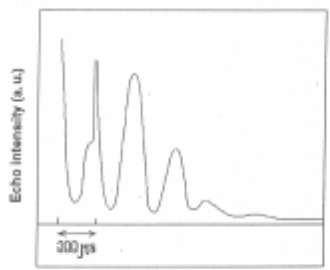
Fig.4a

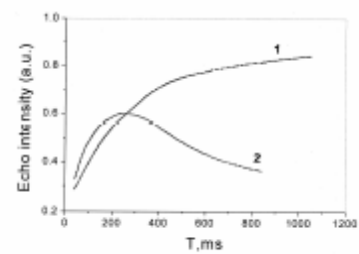
Fig.4b

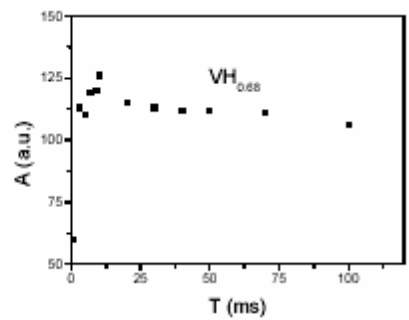
Fig.5